# Mathematical modeling and intuition in microbiology: a perspective

Jamie A. Lopez[1,2,†,*], Amir Erez[3,†,*]
[1] *Department of Bioengineering, Stanford University, Stanford CA 94305, USA*
[2] *Department of Applied Physics, Stanford University, Stanford, CA 94305, USA*
[3] *Racah Institute of Physics, The Hebrew University, Jerusalem 9190401, Israel*
[*] *These authors contributed equally*
[†] *jamie.alc.lopez@stanford.edu*
[†] *amir.erez1@mail.huji.ac.il*

**Abstract**

Mathematical models are increasingly a part of microbiological research. Here, we share our perspective on how modeling advances the discipline by: (i) enforcing logical consistency, (ii) enabling quantitative prediction, (iii) extracting hidden parameters from data, and (iv) generating intuitive understanding. We map a spectrum of modeling frameworks, from whole-cell simulations to minimal logistic growth equations, and provide interactive examples for some common frameworks. Building on this overview, we outline pragmatic criteria for choosing an appropriate level of description to capture phenomena of interest. Finally, we present a case study in modeling of microbial ecosystems from our own work to illustrate how mechanistic modeling can yield generalizable intuition. This perspective aims to be an introductory roadmap for integrating mathematical modeling into experimental microbiology.

**Introduction**

In the physical sciences, theoretical developments can anticipate major experimental observations. Notable examples include the periodic table of elements, gravitational lensing of light, and the Higgs boson. These examples of useful physical theories share many hallmarks, including internal consistency and empirical adequacy, often leading to easy-to-understand and intuitive explanations. Yet in the biological sciences, the role of theory is less defined, though has been discussed for decades [1]. Both the complexity of biological systems and the apparent lack of generalizability between them have made broad theoretical claims difficult to establish. In this perspective, we focus on the subdiscipline of microbiology and outline roles for modeling and how it can be useful. We do not go into detail about particular types of modeling, as there are existing perspectives on the model classes we highlight here [2–4]. Similarly, we use the terms "theory" and "modeling" interchangeably. While some authors make fine-scale distinctions between these two terms, we use both to broadly refer to the building of mathematical representations of the world.

Focusing on microbiology, how are mathematical models created, analyzed, and applied? After sufficient observation of a system, theoretical work typically involves three core steps in a back-and-forth process: (i) formulation: creating an idealized representation of the phenomena of interest, in the context of a specific system. For example, in soil microbiomes, gene expression and taxonomic composition warrant distinct modeling frameworks.; (ii) solution: exploring modeling outcomes via pen-and-paper mathematical analysis and computer simulations, with the hope of learning something about the real system; (iii) validation: applying or testing findings from the model on experimental data. In this perspective, we focus broadly on model formulation and how it is influenced by the intended solution and validation steps.

**The contents of mathematical models in microbiology**

What are mathematical models in microbiology composed of? To predict a single cell's behavior, one approach is to model the entire cell as accurately as possible, as is done in whole-cell computational models. These models attempt to account for every type of molecule and process within a cell, including molecular noise [5]. Working with such models is difficult because of their computational complexity and the large number of parameters needed to be known in advance. However, to capture a handful of behaviors, such as growth rate, antibiotic resistance, or motility, it often suffices to neglect much of the cell's complexity. For example, genome-scale metabolic models (GeMs) still attempt to capture every type of metabolite and enzyme in a cell, but ignore reaction kinetics and molecular noise by assuming that the reaction fluxes in the cell will be tuned to optimize growth [6]. Such models ignoring noise are termed "deterministic", while those incorporating noise are referred to as "stochastic". Despite their simplified structure, GeMs can still accurately predict growth rates and metabolic secretions [7]. Often, a sufficient model can be even simpler than a GeM. In order to predict acetate excretion in *Escherichia coli*, Basan et al. [8] successfully developed a model considering only carbon balance, energy balance, and proteome allocation. Indeed, for many applications, one can entirely neglect cellular processes and consider only thermodynamics. To predict whether a cell can grow in a given condition, one can determine whether that cell's metabolism is thermodynamically favorable, i.e., whether its primary energy-yielding reaction has a negative Gibbs free energy. This approach is formalized in the "redox tower" model of microbial metabolism [9]. A salient example is anaerobic ammonium oxidation (ANNAMOX): at a time when most of the inner workings of cells were unknown, the existence of ANNAMOX was predicted thermodynamically in 1977 by Engelbert Broda [10], (later suspected to be a KGB spy [11]), while the first empirical evidence took almost twenty more years to emerge [12]. Thus, specific aspects of a cell's immense complexity can be well captured using a simple model with judicious assumptions [1]. These models are most effective when they include only those entities and processes essential for capturing the relevant dynamics.

Often in microbiology one wishes to understand not just one cell's behavior, but that of an entire community of microbes [13–15]. This increase in scale creates additional modeling trade-offs: microbial communities often contain millions of cells, so that tracking even a very simple model of every cell becomes impossible. One can still model each cell individually, but only for a small number of cells, as is done in agent-based modeling. For example, to understand the 3D structure of biofilms, Beroz et al. simulated individual growing cells subject to mechanical forces

[16]. Another approach is to neglect individual cells entirely, instead adopting a "continuum" approach where one only tracks the dynamics of continuous variables such as cell density. Conceptually, this means slicing up space into pieces that are large enough to contain many cells so that a meaningful density can be defined, but are small enough that the cells in a given slice are approximately homogeneous. In this vein, the Keller-Segel chemotaxis models can explain the traveling waves of bacteria formed by growing populations of motile chemotactic cells [17]. Similarly, stress-strain considerations explain well the seeding of fruiting bodies in *Myxococcus xanthus* [18]. Very commonly, one can capture behaviors of interest by neglecting spatial structure entirely and modeling only the total population size. These "well-mixed" models are applied widely, ranging from generalized models of ecological interactions such as the Lotka-Volterra variants [19–21] and MacArthur's resource competition model [22–25], to engineering-focused models such as the Anaerobic Digester Model I [26]. Note that the well-mixed assumption does not preclude inclusion of factors such as finite population size or stochasticity.

How does one decide what sort of model to use to capture a phenomenon of interest? One ethos is "models should be as simple as possible, but not more so"[1]. This ethos involves starting with a minimal model and deliberately ignoring as many complexities of the system as possible. Even minimal models can be effective at capturing key features, as microbial systems can exhibit "emergent simplicity" in which microscopic details are not needed for outcome prediction [27]. The minimal approach has the advantage of relatively straightforward analysis and interpretation. Additionally, a simple model formulation does not preclude complex behaviors: the most useful models exhibit behaviors not explicitly included in their formulation. For example, Keymer et al. found that the intricate behavior of *E. coli*'s chemotaxis network can be well-captured by a two-state receptor model with few parameters [28]. The inability to capture key features of the data often compels theorists to increase model complexity, progressively adding parameters and terms to improve empirical fit. Striking the right balance, defining the essential structure and capturing it in the simplest model, is widely regarded as a matter of craft and judgment. It is therefore often prudent to build on established insights, following the maxim: "don't reinvent the wheel". It may be preferable to deploy a more complex modeling framework if it is widely-used in contemporary literature. Using a standard framework shifts the focus from the first part of the model-building process, model formulation, to the second part, modeling results. This approach can be favorable in applied settings, for example in the anaerobic digester models we cited above. The use of standard frameworks facilitates communication across research groups and the application of pre-existing results, though sometimes at the expense of model interpretability. For example, starting with a more complex model may make it more difficult to determine the key processes/components contributing to the phenomenon of interest.

For some problems, there may not be a single 'best' model. Levins, in his seminal analysis of modeling in biology [29], argued that the soundest approach is to treat a biological question with multiple models that share biological assumptions while differing in their simplifying assumptions. Agreement across such models supports a "robust theorem", i.e., a conclusion insensitive to modeling details. Levins memorably described these robust theorems as an

"intersection of independent lies".

Finally, we note that models useful for microbiology can come from other disciplines. For example, models of cancer cell metabolism can provide intuitive insight into regulation of fermentation and respiration in bacteria [30], and models commonly used in microbial ecology often have their origin in broader fields of ecology and population biology [24,29,31,32]. Since models contain only very limited biological detail, it is unsurprising that frameworks prove useful across a wide range of biological contexts.

Before proceeding to discuss model application, we summarize key decisions in choosing a modeling framework, with an eye towards those new to modeling. In [Figure 1A](), we present five simplifications that are commonly considered during the development of a model. A useful strategy is to begin with the most simplifying assumptions possible, then relax these assumptions (e.g., moving from a deterministic model to a stochastic one) as the need arises. In addition to these five modeling decisions, we list common microbiological data types and provide examples of model frameworks that have been used to analyze them in [Figure 1B]().

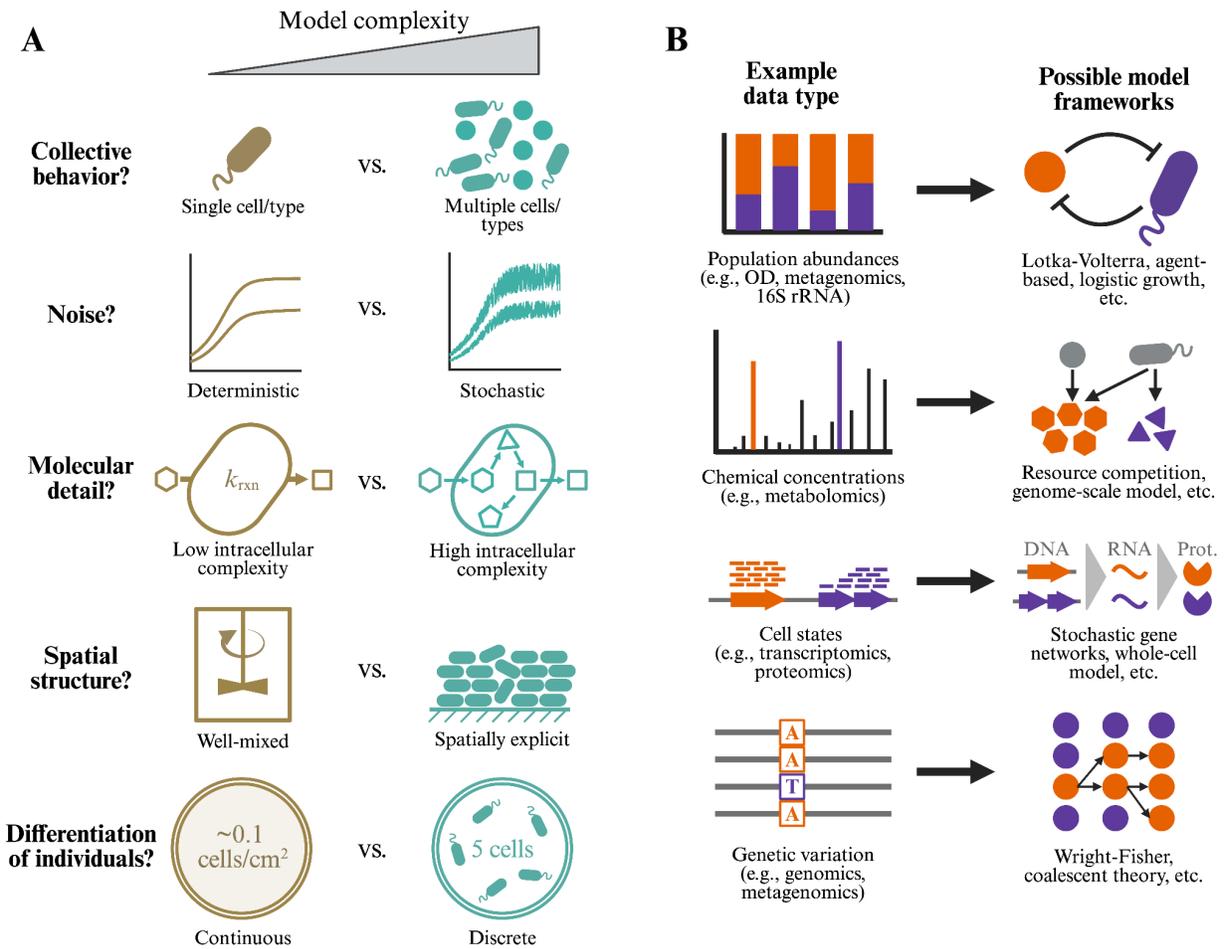

**Figure 1: Overview of model development and choice in microbiology** (this listing is not exhaustive of modeling decisions, data types, or model frameworks). **(A)** Five essential decisions during microbiology model construction. Model complexity increases from left to right, with a continuum of complexity existing between the two extremes shown (e.g., increasingly intricate spatial structure). **(B)** Examples of model frameworks used *in concert* with common microbiological data modalities.

## Applications of mathematical models in microbiology

Once one has selected a model framework, how can the model be useful? We argue model application often falls into one or more of these categories: (i) consistency validation, (ii) quantitative prediction, (iii) data analysis and parameter inference, and (iv) intuition building. We separate inference and prediction as though they both involve estimating model parameters, they differ in their purpose and validation: prediction requires validation on out-of-sample data and therefore demands robust generalization to outcomes not represented in the training set. Furthermore, predictive performance does not require a mechanistic model with interpretable parameters; it may instead rest on empirical associations that persist beyond the data used for fitting. We also distinguish prediction from intuition because they can be competing goals: accurate prediction often requires many data-driven links between model and observations, whereas an abundance of such links can obscure intuitive understanding. For example, traffic is notoriously difficult to model from first principles, but navigation software uses machine learning to predict traffic remarkably well without being able to explain it.

Note that in practice, one often goes back and forth between model application and framework choice until converging on a suitable model. Below, we provide examples of useful model applications though the lens of our proposed framework:

**i. Consistency.** The act of constructing a model serves to formalize assumptions and hypotheses in a mathematical framework. Even before assumptions can be made, biological quantities of interest must be defined, which can be difficult. This is highly beneficial, as these quantities may be otherwise left vaguely defined. The act of modeling itself also confers another immediate benefit: the model must be internally consistent. While trained readers may sometimes detect fallacies or contradictions in verbal arguments, translating these arguments into mathematical form, with well-defined quantities, subjects them to the stricter and more transparent constraints of formal logic. An important part of hypothesis-driven science is conditional reasoning of the form "if X, then Y". However, when working with a complex biological system, determining the implications of a set of assumptions (e.g., the presence of a feedback loop) can be difficult. A model allows one to rigorously understand the consequences of a set of assumptions. A notable example of this can be found in the work of Robert May, who sought to understand the relationship between ecosystem complexity and ecosystem stability, where complexity is defined by the strength of inter-species interactions and the number of species in the ecosystem. Before May, it was thought that the more complex an ecosystem is, the more stable it would be. This intuition stems from many observations that when enough species are removed from an ecosystem, it undergoes abrupt and major shifts. However, in a groundbreaking use of modeling, such vague intuition could be probed by a specific and definitive calculation. Using random matrix theory in conjunction with traditional ecological models of species interactions [31], May showed that increased complexity would in fact lead to a loss of stability [33]. May's stability criterion is neatly summed up by the expression $\sigma\sqrt{nC} < 1$ where $\sigma$ is the typical strength of interactions, $n$ is the number of species in the ecosystem and $C$ is the fraction of species pairs that interact directly. Thus, May's application of modeling allowed for a more rigorous posing of hypotheses, inspiring a wide range of future work [20,34–

**Ii. Prediction.** Another application for models is quantitative prediction: attempting to predict the behavior of an experimental system. For example, the aforementioned models of anaerobic digesters have been used successfully in industrial practice [26]. More recently, Sun et al. used a mechanistic model of microbial denitrification to quantitatively predict rates of biogeochemical reactions in the ocean [38]. Note that this quantitative application of mathematical modeling, particularly in biology, can be difficult. Models suitable for quantitative predictions can be very complex, requiring a great deal of experiments to parameterize. Indeed, if quantitative prediction is the primary goal, then more data-driven approaches, such as machine learning [39], can be superior. However, though very successful in many ways, machine learning does not replace mechanistic modeling when trying to produce intuitive explanations. Indeed, machine learning can often be optimized when guided or constrained by principles uncovered through mechanistic analyses. Similarly, machine learning analyses can be useful to guide the development of mechanistic models, identifying statistical relationships that point towards underlying mechanistic processes.

36]. We provide interactive Python simulations of both the classical Lotka-Volterra dynamics and May's stability criterion in this manuscript's code repository [37].

**Iii. Analysis and inference.** Modeling can also be used to interpret and gather information from existing data. For instance, one may infer parameters of interest using nontrivial manipulations of existing data. Suttle et al. [40] used simple mathematical models of microbe-phage dynamics in conjunction with existing measurements to infer that phage are a major driver of surface ocean ecosystems, a calculation that strongly informs our current understanding of surface ocean waters [41]. In a more recent example, Mulla and Müller et al. used a mathematical model of bacteriophage dynamics to enable high-throughput inference of phage growth and parameters [42]. Models can also enable more qualitative analysis of experimental data. Famously, Luria and Delbrück used a model to interpret results from an experiment in which the number of bacteria surviving a bacteriophage challenge is measured. Their goal was to determine if bacterial phage resistance mutations occur independently or only in response to phage selection pressure. The key contribution of their model was not to quantitatively interpret the results of their experiments, but rather to define null and alternative hypotheses: if mutations arise independent of selection pressure, the variance of an experimental outcome would be far higher than its mean [43]. This allowed them to interpret the observed large inter-experiment variance as a signal of independent mutations. Beyond this qualitative analysis, the model developed by Luria and Delbrück was used to develop methods and experimental frameworks to estimate microbial mutation rates.

**iv. Intuition.** Finally, modeling can be used to build intuition about aspects of microbiological systems. We define "intuition" in this case as mental heuristics that enable one to better navigate and study biological systems. Often, intuition points one towards the important processes and parameters in a system, reducing complexity. The boundaries between intuition-building and the three preceding uses of modeling are often indistinct. For example, in the Luria-Delbrück experiment discussed above, the model provided generalizable intuition in addition to outlining null and alternative experimental outcomes. Luria and Delbrück found that the high variance in their independent mutation model arises from "jackpots", rare cases where a mutation occurs very early in the experiment and produces a massive population of mutant cells. This intuition of "jackpot" events being a major driver of variation in evolutionary dynamics has found wide application, leading to a general class of experimental "fluctuation tests" [44] and advances in fields such as population genetics [45]. A more recent example of intuition-building theory comes from the work of Hwa and colleagues [8,46]. They set out to develop biophysical laws governing cellular growth, successfully developing models useful for quantitative prediction of phenomena such as overflow metabolism. But, just as important as their predictive ability, these models provide intuition that cell growth can be independent of many molecular details, instead being captured by coarse-grained quantities such as proteome allocation and the ratio of cellular RNA to protein. This simplifying intuition enables the design of more parsimonious and efficient experiments and models. Being able to think about cells in such broad terms while remaining close to experimental reality is conceptually similar to how a gas can be described by its temperature, pressure and so forth, without resorting to a microscopic description of every gas molecule, or even, the electrons, protons, and quarks inside it. Obviously, such approaches are astonishingly successful in describing "dead" physical matter, and Hwa's work is exemplary of the current ongoing effort to similarly succeed in describing living matter.

Our classification of model applications is not the only possible framework. Levins [29] suggested a different approach centered around maximization of three goals: generality, realism, and precision. Levins argues that in practice one can choose only two of the three goals and that different modeling approaches trade among these outcomes, depending on whether the immediate objective is understanding, prediction, or intervention. For example, he contrasts approaches that prioritize realism and precise prediction in a narrow domain with those that prioritize generality and analytic tractability (often via unrealistic assumptions). While Levins' framework is not directly equivalent to the one we propose, there is a rough mapping between them, e.g., sacrificing generality while maintaining realism and precision will give a model that can accurately predict a system in a narrow range of conditions. Despite differences in formulation, both our frameworks share foundational assumptions. A key assumption is that biological systems contain far more mechanistic degrees of freedom than can be feasibly measured. Strikingly, this underdetermination remains a defining feature of biological inquiry sixty years after Levins' manuscript, even with dramatic improvement in measurement technologies.

**An in-depth personal example of intuition-building theory in microbiology**

As an example of how simple intuitive understanding emerges from modeling, we provide a story from our own research in microbial ecology. The story begins with a relatively simple model that exhibited complex behavior, which we analyzed using a suite of mathematical approaches. After being encouraged by a referee to seek an intuitive explanation of our modeling results, we realized that the complex behaviors we observed could be explained in a simple and more useful way than our formal mathematical results.

We were motivated by a substantial gap between existing modeling and experimental approaches. We modeled an ecosystem using dynamic variables for the species abundances and for the nutrient and metabolite levels, embodied by the classical MacArthur consumer-resource models [22,47–49]. For decades, the dominant form of these models studied by theoretical ecologists has been the chemostat [50]: a well-mixed ecosystem into which nutrients continuously flow, with a corresponding constant removal of both biomass and metabolites [51–54]. However, chemostats are difficult to implement experimentally, requiring great care to maintain steady conditions. Much more common experimentally is the serial dilution ecosystem, in which nutrient addition and biomass removal is performed periodically, typically once every 24-48 hours (this time is generally adjusted to encompass multiple generations) [27,55–57]. Many natural ecosystems feature periodic fluctuations that can be approximated by serial dilution, including the dynamics of healthy intestines and day-night cycles [58,59], oceanic tides [60], and the seasons of the year [61,62]. Little theory had been developed for these serial dilution ecosystems, as they are much more difficult to analyze mathematically than their chemostat counterparts: chemostat models reach a stable steady state, whereas the "steady-state" of a serial dilution ecosystem is a complex species abundance time course that repeats from batch to batch (Figure 2A). This difference between theoretical and experimental approaches led to an apparent disconnect, whereby much of the theory invoked to analyze serial dilution experiments did not account for the periodic fluctuations that dominate these systems [27,55,56,63,64].

We set out to extend an existing chemostat theory [52] to serial dilution, validating that the nutrient consumption and growth dynamics we assumed were quantitatively consistent with a prior serial dilution experiment. This modeling framework simplified reality to consider the serial dilution ecosystem as defined by the species that it contains and each species' metabolic strategy (Box 1). The metabolic strategy defines how much metabolic capacity a species dedicates to each nutrient, assuming metabolic capacity has a fixed budget. The model displayed a wide variety of emergent, counter-intuitive states that do not occur in chemostat models (Figure 2B). For example, we found situations where the model would exhibit high species diversity in nutrient-poor conditions, but underwent a collapse in diversity when more nutrients were provided. Paradoxically, when even more nutrients were provided, the community diversity would recover. Even in the absence of these major collapses, community composition was sensitive to nutrient levels, and the directions of these trends varied widely [49,62,65]. The periodic fluctuations characteristic of serial dilution had a major impact on the ecology and should be considered when analyzing and modeling experimental data. But how to turn these theoretical observations into something useful for experimentalists?

First, we applied a standard dynamical systems toolkit to assess how ecosystem composition responds to nutrient supply. This formal analysis revealed the sensitivity of community structure to nutrient levels and provided a mechanistic explanation for the observed diversity shifts. Having submitted our work for publication, the reviews came back with a major demand: more intuition. The request for intuitive understanding did not surprise us, as it was not immediately obvious how to translate our results into usable insights about microbial ecosystems. A brainstorming session one afternoon led us to realize that the complicated behaviors we had observed in the serial dilution model could be explained by a single intuitive concept, which we called the "early-bird" effect ([Figure 2C](#)).

Essentially, in serial dilution, early advantages can compound. If an organism is initially able to grow rapidly, it can later leverage that larger population size to outcompete otherwise superior competitors. It gains advantage by depriving competitors of nutrients, despite being less effective per capita at consuming the remaining nutrients, due to its increased population. The larger population caused by the fast, early growth compensates for a growth disadvantage later in the batch. This differs from a classical priority effect [66]. The priority effect, in contrast to the early-bird effect, reflects history-dependent advantages from arrival order across community assembly. The early-bird effect instead arises dynamically within each serial-dilution batch from asymmetric initial resource pulses that give specialists a compounding lead. The early-bird effect becomes more powerful the more nutrient is provided, but can eventually be limited by saturation of growth rates, thus explaining the non-monotonic dependency on nutrient levels. The early-bird explanation proved more useful to our experimentalist colleagues. If one could identify the initial fast grower in the ecosystem, one can predict how community composition will change when nutrient levels shift. Our theory was later validated in gut-derived microbial communities relying almost entirely on the simple intuition of the early-bird effect [67]. Later work on antimicrobial resistance and microbial debt [68] was guided by the organizing principle of the early-bird effect. The intuitive explanation became so prominent in our minds that though it was added at the very final stage of a multi-year project, we now find ourselves referring to the manuscript as "the early-bird paper".

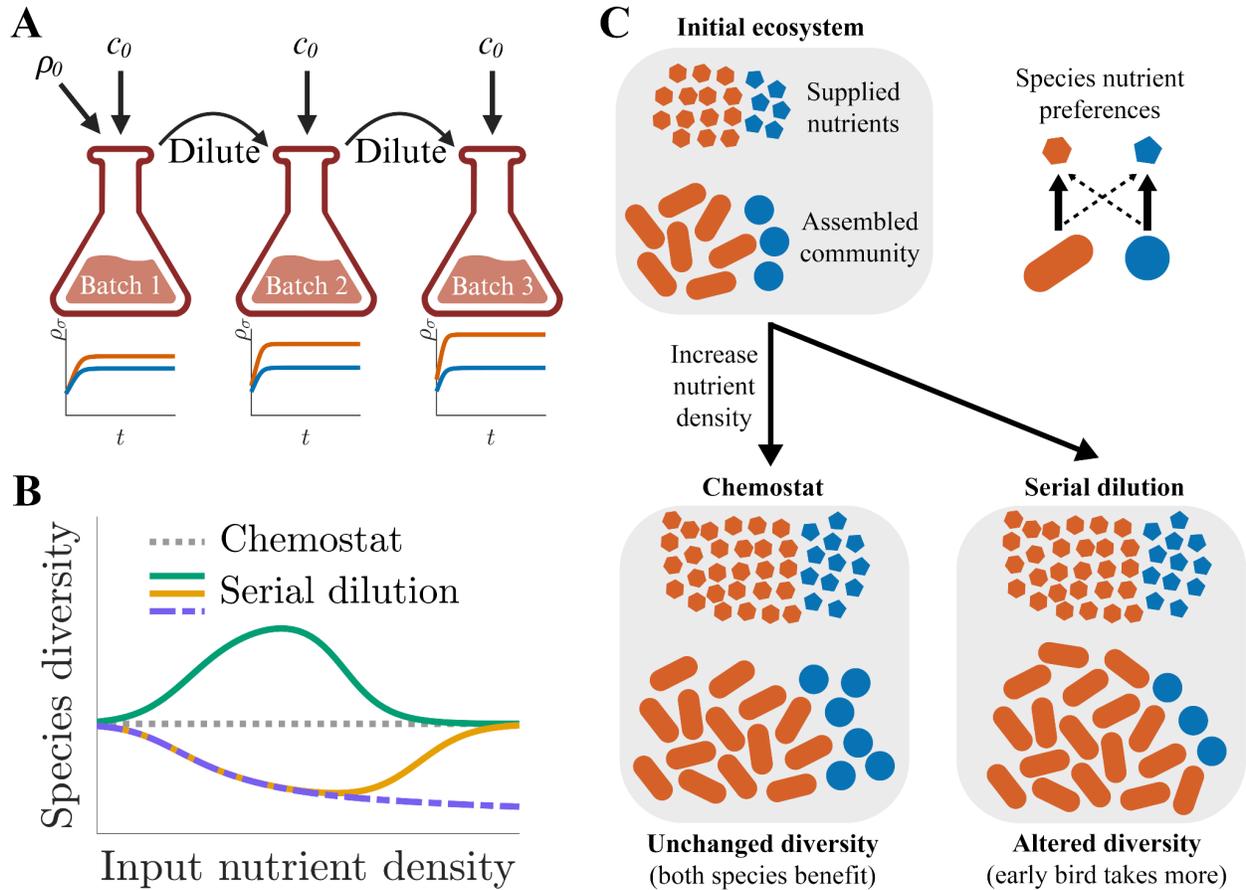

**Figure 2: Serial dilution and the early-bird effect: an in-depth example of mathematical modeling in microbiology. (A)** Schematic of serial dilution model. In these ecosystems, an initial bacterial population $\rho_0$ and initial nutrient amount $c_0$ are added to a well-mixed ecosystem. After a set time, the bacteria are diluted into a new ecosystem ("batch") and the nutrients are replenished. This process continues for a given number of batches. **(B)** Relationship between diversity and supplied nutrient in the chemostat and serial dilution models. In the chemostat model, there is no change in diversity when supply rates are changed. In contrast, different serial dilution systems can exhibit a wide range of behaviors, from decreasing to the "hump-shaped", mirroring the diversity seen in real ecosystems [41]. **(C)** Intuition for serial dilution ecosystem behaviors. Many differences between the chemostat and serial dilution can be explained intuitively by the "early-bird" effect. In serial dilution, the time lag between growth and dilution allows species that grow early in the batch to take advantage of their larger population size. As a result, they can exploit more of nutrients that they are otherwise poorer competitors for. This effect changes in strength with nutrient level, explaining the observed nutrient-dependent effects.

## Summary

While the role of modeling in microbiology is less established than in the physical sciences, there is already a burgeoning set of examples and practices of how to apply models in the field. To show how models can benefit many stages of research, we have presented a variety of seminal contributions of modeling to microbiology, alongside a more in-depth narrative from our own work. Before performing an experiment, a model forces us to define the underlying assumptions, and can provide null expectations informing study design. More quantitative models can provide predictions, enabling bioengineering and bioprocess development. After experiments are performed, models can help us better interpret the massive quantities of data generated by modern microbiology, and be used to extract novel insights from existing data. Finally, models can provide intuition that enhances our understanding of the microbiological world and unifies disparate concepts. At its best, modeling is fun as well as useful. Akin to how children discover the world by playing pretend, exploration of the hypothetical worlds contained in mathematical models allows intuition about complex systems to quietly take root.

## Acknowledgements

We thank the following people for useful suggestions: Nathalie Balaban, Omri Finkel, Jonathan Friedman, Jacopo Grilli, Bruce Levin, Yigal Meir, Sara Mitri, Avaneesh Narla, Nasa Sinnott-Armstrong, Xin Sun, Sophie Walton, Ned Wingreen, and Katherine Xue.

## Conflict of interest

The authors declare no conflicts of interest.

## Data Availability

This manuscript contains no data. The example code can be found in
https://github.com/AmirErez/Manuscript-TheoryPerspective

**Box 1: Tutorial. Consumer-resource dynamics in microbial ecosystems**

We have discussed mathematical models of microbial systems in general terms. For completeness, we now provide a detailed example and interactive simulation of one such model. This section is more technical than the rest of the manuscript, and together with the code examples provided, can serve as a springboard for those interested in modeling. Consumer-resource models describe the dynamics of both microbial species (consumers) and nutrients (resources), offering insights into ecosystem stability and diversity [22,47–49]. We extended these models to describe serial-dilution cultures, systems where communities grow in discrete nutrient pulses separated by dilution events ([Figure 2](#)) [49,62,65,68]. This serial-dilution model mimics not only common microbiological experiments, but is also a simple representation of natural fluctuations. At the start of each batch, a set of species are used to inoculate the culture, together with a nutrient bolus. We focus on limiting substitutable nutrients (e.g., two sources of carbon that are the bottleneck for microbial growth). In this box, we give an example with two species, A and B, whose concentrations are denoted $\rho_A(t)$ and $\rho_B(t)$, and two nutrients types, $c_1(t), c_2(t)$. Note that time $t = 0$ indicates the beginning of a batch. Growth proceeds for a specified time ($t = t_f$), usually until nearly all nutrients are depleted. Assuming complete depletion, the total biomass at the end of a batch will be the sum of the initial biomass and the supplied nutrients nutrients: $\rho_A(t_f) + \rho_B(t_f) = \rho_A(0) + \rho_B 0) + c_1(0) + c_2(0)$. For simplicity, we assume a one-to-one biomass-to-nutrient yield for both species. The final community composition at time $t_f$ sets the inoculum composition for the subsequent batch. This growth-dilution process repeats iteratively until the ecosystem reaches steady-state, which occurs when the relative species abundances at time 0 are the same as at time $t_f$. Indeed, in this steady state of the batch culture, all surviving species grow by the same fold change during each batch. Each species has a defined enzyme strategy, i.e., for species A, $(\alpha_{A,1}, \alpha_{A,2})$ and similarly for species B, with each $\alpha$ determining its maximal nutrient consumption rates (and therefore maximal growth) rates on each nutrient. Nutrient consumption rates are described by classical Monod kinetics, e.g., for species A and nutrient 1, the Monod consumption is $j_{A,1}(t) = \alpha_{A,1} \frac{c_1(t)}{c_1(t) + K}$, which reaches its maximum, $\alpha_{A,1}$, at saturating nutrient concentrations when $c_1 \gg K$, with $K$ is a half saturation constant we simplify as equal for all species and nutrients. The dynamics within a batch are thus:

$$\frac{dc_1}{dt} = -\rho_A j_{A,1} - \rho_B j_{B,1} \; , \; \frac{dc_2}{dt} = -\rho_A j_{A,2} - \rho_B j_{B,2} \, , \; \frac{d\rho_A}{dt} = \rho_A(j_{A,1} + j_{A,2}), \frac{d\rho_B}{dt} = \rho_B(j_{B,1} + j_{B,2}) \; .$$

Though in general these equations cannot be solved analytically, simulating these dynamics numerically can be readily done in most programming languages. We have created an interactive Jupyter notebook that allows one to simulate these equations in a variety of parameter ranges (supporting information). One can modify the model to incorporate more realistic dynamics, such as more species and nutrient types, byproduct production and cross-feeding. We invite the reader to explore these dynamics.

**Mutual invasibility and species coexistence:**
A powerful criterion for coexistence is mutual invasibility [70]. This involves examining the growth dynamics when a community containing only one resident species at steady state is perturbed by introducing a tiny quantity of another (invader) species. Concretely, one assumes only species A exists, controlling $c_1(t)$ and $c_2(t)$. A tiny amount of an invader species would either increase or decrease given these nutrient dynamics. Coexistence requires that each species can invade and grow when rare against the other. This condition can predict stable coexistence versus competitive exclusion outcomes. By using this mutual invasibility condition we were able to infer the coexistence region in the serial dilution regime. We invite the reader to use the above simulation to explore the concept of mutual invasibility.